\documentclass[aps,prb,twocolumn,groupedaddress,showpacs]{revtex4}
\usepackage{graphicx,amssymb,amsbsy,color}
\input epsf % defines \epsfbox and supporting macros
\epsfverbosetrue % messages will show height and width

\begin{document}

\title{Oscillations of Bose condensates in a one-dimensional optical superlattice}
\author{Chou-Chun Huang and Wen-Chin Wu}
\affiliation{Department of Physics, National Taiwan Normal
Univesity, Taipei 11650, Taiwan}

\begin{abstract}
Oscillations of atomic Bose-Einstein condensates in a 1D optical
lattice with a two-point basis is investigated. In the
low-frequency regime, four branches of modes are resolved, that
correspond to the transverse in-phase and out-of-phase breathing
modes, and the longitudinal acoustic and optical phonon modes of
the condensates. Dispersions of these modes depend intimately on
the values of two intersite Josephson tunneling strengths, $J_1$
and $J_2$, and the on-site repulsion $U$ between the atoms.
Observation of these mode dispersions is thus a direct way to
access them.
\end{abstract}

\maketitle

Cold atoms in optical lattices is currently an active research
topics in the cross-disciplined field of atomic molecular and
optics physics and condensed-matter physics. From the condensed
matter point of view, because (i) laser power can be tuned to vary
the ratio of on-site repulsion to inter-site coupling of cold
atoms, (ii) different geometry of laser beams can be taken to
manipulate different dimensions and configurations of optical
lattices, (iii) different types of atoms (fermions, bosons, ions,
or their mixtures) can be loaded into the optical lattice,
consequently the simulation of condensed matter environment is
considerably easy and flexible. Recent investigations on the
cold-atom system with an optical lattice have been made in various
aspects, including the superfluid (SF)-Mott insulator quantum
phase transition \cite{jaksch98,greiner02}, the band-structure
phenomena \cite{machholm03,wu00}, and the quantum informatics
\cite{cirac03,vollbrecht04}.

\begin{figure}[th]
\vspace{-2.4cm} \mbox{\epsfxsize=1.2\hsize{\epsfbox{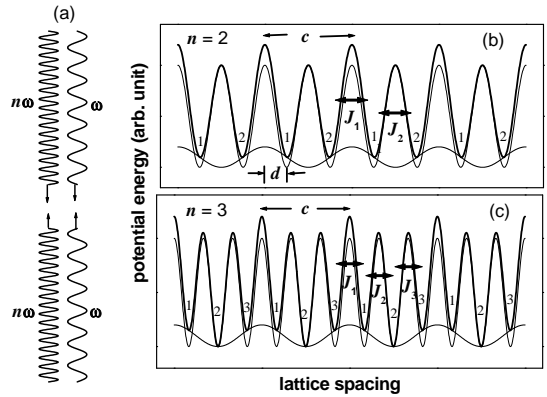}}}
\vspace{-7.6cm} \caption {(a) Illustration of how a 1D optical
lattice with an $n$-point basis is formed by two laser beams and
their reflected beams. Frame (b) and (c) show the resulting
potential energies (thick solid lines) as a function of lattice
spacing of the $n=2$ and $n=3$ case. $c$ denotes the lattice
constant and $J$'s denote the Josephson tunnelling couplings. In
frame (b), $d$ denotes the spacing between any potential maximum
and its nearby minimum.} \label{fig1}
\end{figure}

In this paper, we consider a physical environment very common in
crystalline solids. It is often in a solid that the system is best
described by a superlattice ({\em i.e.}, lattice with a basis).
This occurs when there are more than one type of atoms (or sites)
within a unit cell. Correspondingly in the cold-atom system, a 1D
optical lattice with $n$-point basis can be set up as follows. One
can shine two laser beams coherently towards the same direction
(say, $z$-axis), with one having $n$ times of frequency (and thus
$n$ times of wave vector) of the other. Along with their reflected
beams, the resulting electric field is
$E(z,t)=A_1\cos(nkz)e^{in\omega t}+A_2\cos(kz)e^{i\omega t}+{\rm
c.c.}$ and the dipole energy, which is proportional to the time
average of electric field square, is thus
\begin{eqnarray}
U(z)\propto\overline{E(z,t)^2}\sim
A_1^2\cos^2(nkz)+A_2^2\cos^2(kz). \label{eq:trap}
\end{eqnarray}
The cross term vanishes due to time average. Consequently Bose
condensates, which is pre-confined in the $x$-$y$ plane by the
magnetic trap, can be optically trapped and form a $z$-direction
1D optical lattice with an $n$-point basis. Since the ratio of
amplitude $A_1/A_2$ can be adjusted and the factor $n$ can be
chosen, 1D optical superlattice can be created for a great
flexibility. Figure~\ref{fig1} illustrates how the 1D optical
superlattice with a 2 and 3-point basis is formed and their
resulting potential energies ($A_1/A_2=\sqrt{5}$ is taken).
Josephson coupling $J$'s that differ due to condensates tunnelling
across different potential barriers are also indicated.

Since the achievement of atomic BEC in 1995, dynamics of single
Bose condensate had been under detailed investigation both
theoretically and experimentally \cite{dalfovo99}. When
condensates are distributed in an optical lattice, in addition to
collective modes, phonons can also propagate among the lattice.
How the collective and phonon modes are coupled in such a
condensed system thus has attracted a large number of studies
\cite{kramer02,berg98,javanainen99,machholm03,taylor03,martikainen04}.
In fact, condensate excitations in a 3D optical lattice have been
measured recently in the SF regime \cite{schori04}. Among
different theoretical approaches, Martikainen and Stoof
\cite{martikainen04} have recently applied the variational
approach to study the transverse breathing mode and longitudinal
acoustic phonon in a simple 1D Bose-condensed optical lattice. The
advantage of this approach is that it enables naturally the
coupling between the breathing and phonon modes, and at the same
time, allows for analytical results. In the present context, we
shall generalize their approach to study how the breathing and
phonon modes of Bose condensates are coupled and behave in a 1D
optical superlattice.
%In order to save the space, we have used the same
%notations as those in Ref.~\cite{martikainen04} without giving
%further definition. So readers should refer to
%Ref.~\cite{martikainen04} for more detailed ....................

At sufficiently low temperatures, dynamics of Bose condensates is
governed by the time-dependent Gross-Pitaveskii (GP) energy
functional
\begin{eqnarray} E\left[\Psi^*,\Psi\right]&=&\int d^3r
\left\{-{\hbar^2\over 2m}\Psi^*\nabla^2\Psi+\left[V_{\rm
trap}({\bf r})\right.\right.\nonumber\\ &+&\left.\left.{U_0\over
2}|\Psi|^2-\mu\right]|\Psi|^2\right\}, \label{eq:GP}
\end{eqnarray}
where $\Psi=\Psi({\bf r},t)$ is the time-dependent wave function
of the condensate, $\mu$ is the chemical potential, and
$U_0=4\pi\hbar^2 a/m$ with $a$ the $s$-wave scattering length of
the two-body interaction. In the present case of interest, the
external trap potential has two contributions, $V_{\rm trap}({\bf
r})=V_m({\bf r})+V_o({\bf r})$, where $V_m({\bf r})={1\over
2}m[\omega_\rho^2 \rho^2+\omega_z^2 z^2]$ with $\rho^2\equiv
x^2+y^2$ is due to the magnetic trap, while $V_o({\bf
r})=V_o(z)=V_1 \cos^2\left({2\pi z\over c} \right)+V_2
\cos^2\left({\pi z\over c} \right)$ is due to the optical trap.
When $\omega_\rho\gg\omega_z$ for the magnetic trap, optical
potential then results a 1D optical superlattice (along $z$ axis)
with a two-point basis. Here $c$ corresponds to the lattice
constant which is related to the wavelength ($\lambda$) or the
wavevector ($k$) of the laser beam with frequency $\omega$ by
$c=\lambda/2=\pi/k$ (see Fig.~\ref{fig1}). It is convenient to
work with dimensionless quantities and rescale the energy, time,
and length as $E/(\hbar\omega_\rho)\rightarrow E$,
$t\omega_\rho\rightarrow t$, and ${\bf r}/\ell_\rho\rightarrow
{\bf r}$ ($\ell_\rho \equiv \sqrt{\hbar/m\omega_\rho}$).
Eq.~(\ref{eq:GP}) thus becomes
\begin{eqnarray} E\left[\Psi^*,\Psi\right]&=&\int d^3r
\left\{-{1\over 2}\Psi^*\nabla^2\Psi+\left[{1\over
2}(x^2+y^2)\right.\right.\nonumber\\ &+&\left.\left. \bar{V}_o(z)
+{\bar{U}_0\over 2}|\Psi|^2-\bar \mu\right]|\Psi|^2\right\},
\label{eq:GP2}
\end{eqnarray}
where the dimensionless coupling $\bar{V}_o(z)\equiv
V_o(z)/(\hbar\omega_\rho)$, $\bar{U}_0\equiv 4\pi a/\ell_\rho$,
and $\bar\mu\equiv \mu/(\hbar\omega_\rho)$. The frequency
$\omega_z$ is completely dropped in (\ref{eq:GP2}).

For longitudinal phonon modes to be considered under the {\em
harmonic} approximation, {\em tight-binding} (TB) limit is
applied. This means that anharmonic effect is negligibly small.
This also means that optical potential barriers need to be large,
so condensate wave functions are strongly localized in
$z$-direction around the potential minimums. In this limit,
$\Psi({\bf r},t)$ can be taken to be the form
\begin{eqnarray}
\Psi({\bf r},t)&=&\sum_\ell\left\{w_1[z-\ell
c-d]\Phi_1(x,y,\ell;t)\right.\nonumber\\
&+&\left.w_2[z-(\ell+1)c+d]\Phi_2(x,y,\ell;t)\right\},
\label{eq:wannier}
\end{eqnarray}
where $\ell$ sums over all unit cells and $d$ is the spacing
between any potential maximum and its nearby minimum (see
Fig.~\ref{fig1}). Pertaining to site $1$ and site $2$, $w_1$ and
$w_2$ are the strongly localized Wannier functions depending on
$z$ only, while $\Phi_1$ and $\Phi_2$ are the ones associated with
$x,y$ coordinates which in turn contain the periodic factor
$\sim\exp(i\ell kc)$ for any propagating wave (along the lattice
direction) of wave vector $k$. It is noted that $w_1(z)=w_2(-z)$
for the reflection symmetry. Moreover, for simplicity,
fluctuations (time dependence) are considered to be through
$\Phi_i(x,y,\ell;t)$ only.

Substitution of (\ref{eq:wannier}) reduces (\ref{eq:GP2}) into
\begin{eqnarray}
E&=&\sum_\ell\sum_{i=1,2}\int d^2\rho \left\{-{1\over
2}\Phi^*_i(\ell)\nabla^2\Phi_i(\ell)\right.\nonumber\\
&+&\left.\left[{1\over 2}\rho^2+{U\over 2}
|\Phi_i(\ell)|^2-\bar\mu\right]|\Phi_i(\ell)|^2\right\}\nonumber\\
&+&\sum_\ell\int d^2\rho\left[J_1\Phi_1^*(\ell)\Phi_2(\ell-1)+{\rm
c.c.}\right] \nonumber\\ &+&\sum_\ell\int d^2\rho
\left[J_2\Phi_1^*(\ell)\Phi_2(\ell)+{\rm c.c.}\right],
\label{eq:E1}
\end{eqnarray}
where $\Phi_i(\ell)\equiv \Phi_i(x,y,\ell;t)$ and
\begin{eqnarray}
J_1&\equiv&-\int dz~ w_1^*(z-d)\left[-{1\over
2}{\partial^2\over\partial z^2}+\bar{V}_o(z)\right]w_2(z+d)\nonumber\\
J_2&\equiv&-\int dz~ w_1^*(z-d)\left[-{1\over
2}{\partial^2\over\partial z^2}+\bar{V}_o(z)\right]w_2(z-c+d)\nonumber\\
U&\equiv& \bar{U}_0\int dz~ |w_1(z)|^4=\bar{U}_0\int dz~
|w_2(z)|^4. \label{eq:JJu}
\end{eqnarray}
In (\ref{eq:E1}) and (\ref{eq:JJu}), only nearest-neighbor
intersite couplings are considered. $J_1$ and $J_2$ are thus the
two {\em different} nearest-neighbor Josephson couplings
responsible for condensate tunneling. In principle, $w_1,w_2$ can
be solved numerically, which in turn solve $J_1$, $J_2$, and $U$.
Typical value of $J_1$ or $J_2$ is less than $0.1$ in the TB
limit.

In the following, as mentioned before, we will focus on the
breathing and phonon modes. Similar to Martikainen and Stoof
\cite{martikainen04}, a Gaussian ansatz
\begin{eqnarray}
&&\Phi_i(x,y,\ell;t)=\sqrt{N[1+\delta_{\ell,i}(t)]
B_0[1+\epsilon_{\ell,i}^\prime(t)]\over\pi}\label{eq:Phi1}\\
&\times&\exp{\left[-{B_0[1+\epsilon^\prime_{\ell,i}(t)+
i\epsilon^{\prime\prime}_{\ell,i}(t)](x^2+y^2)\over
2}+i\nu_{\ell,i}(t)\right]}\nonumber
\end{eqnarray}
is assumed for the wave functions, where $N$ represents the
average number of atoms per site and $B_0$ represents the
condensate size in the equilibrium state. In fact, when $N$ is
fixed, the value of $B_0$ can be calculated through minimizing the
GP energy in (\ref{eq:E1}). This was done in
Ref.~\cite{martikainen04} where $B_0=1/\sqrt{1+2U^\prime}$ with
$U^\prime\equiv 4NU/\pi$ is given. No cross ($xy$) term is
considered in (\ref{eq:Phi1}) because breathing mode is isotropic
in the $x$-$y$ plane. For each site $(\ell,i)$, dimensionless
dynamical variable $\epsilon_{\ell,i}^\prime(t)$,
$\epsilon_{\ell,i}^{\prime\prime}(t)$,  $\delta_{\ell,i}(t)$, and
$\nu_{\ell,i}(t)$ corresponds respectively to fluctuations of the
{\em local amplitude},  the {\em local phase}, the {\em number} of
atoms, and the {\em global phase} of the condensate.

One crucial aspect on (\ref{eq:Phi1}) is however that it enables
naturally the coupling between the transverse breathing and
longitudinal phonon modes~\cite{martikainen04}. In a single
Bose-condensed system with repulsive interaction, the fluctuation
of condensed atom number is coupled to the fluctuation of
condensate size. When condensates are distributed in an optical
lattice, the fluctuating degrees of freedom are complicated by the
intersite coupling of Josephson tunnelling. For the present
optical superlattice, the case is further complicated by the
possible out-of-phase motion in addition to usual in-phase one.
How important are the various couplings and how the condensates
are fluctuating in such a system are thus the subject of the
following studies.

By variational approach, one starts from the Lagrangian of the
system, $L=T-E$. Here $T=\int d{\bf r}~{i\hbar\over
2}(\Psi^*\partial \Psi/\partial t-\Psi\partial \Psi^*/ \partial
t)$ and $E$ is given by (\ref{eq:GP2}) [and hence (\ref{eq:E1})].
Applying (\ref{eq:Phi1}) in $T$ and $E$ and expanding to second
order in dynamical variables, one obtains
\begin{eqnarray}
&&{T\over
N}\label{eq:T}=\sum_\ell\sum_{i=1,2}\left\{(1+\delta_{\ell,i})\left[
-\dot{\nu}_{\ell,i}+{\dot{\epsilon}^{\prime\prime}_{\ell,i}\over
2}\left(1-\epsilon_{\ell,i}^\prime\right)
\right]\right\},\nonumber\\
\end{eqnarray}
and
\begin{eqnarray}
&&{E\over
N}=\sum_\ell\sum_{i=1,2}\left\{(1+\delta_{\ell,i})\left[{1\over
2B_0}\left(2+(\epsilon_{\ell,i}^\prime)^2\right)\right.\right.\label{eq:E2}\\
&&\left.\left.+{B_0\over
2}(\epsilon_{\ell,i}^{\prime\prime})^2+U^\prime
B_0\left(1+\epsilon_{\ell,i}^\prime\right)
\delta_{\ell,i}-\bar{\mu}\right]\right\} +E_J, \nonumber
\end{eqnarray}
where the Josephson-tunneling term
\begin{eqnarray}
&&E_J={J_1\over 4}
\sum_\ell[(\epsilon^\prime_{\ell,1}-\epsilon^\prime_{\ell-1,2})^2
+2(\epsilon^{\prime\prime}_{\ell,1}-\epsilon^{\prime\prime}_{\ell-1,2})^2\nonumber\\
&&+4(\nu_{\ell,1}-\nu_{\ell-1,2})^2
+(\delta_{\ell,1}-\delta_{\ell-1,2})^2\nonumber\\&&-
4(\nu_{\ell,1}-\nu_{\ell-1,2})(\epsilon^{\prime\prime}_{\ell,1}
-\epsilon^{\prime\prime}_{\ell-1,2})-4(\delta_{\ell,1}+\delta_{\ell-1,2})]\nonumber\\
&&+{J_2\over 4}
\sum_\ell[(\epsilon^\prime_{\ell,1}-\epsilon^\prime_{\ell,2})^2
+2(\epsilon^{\prime\prime}_{\ell,1}-\epsilon^{\prime\prime}_{\ell,2})^2\nonumber\\
&&+4(\nu_{\ell,1}-\nu_{\ell,2})^2+(\delta_{\ell,1}-\delta_{\ell,2})^2\nonumber\\&&-
4(\nu_{\ell,1}-\nu_{\ell,2})(\epsilon^{\prime\prime}_{\ell,1}
-\epsilon^{\prime\prime}_{\ell,2})-4(\delta_{\ell,1}+\delta_{\ell,2})].\label{eq:EJ}
\end{eqnarray}
It is noted in (\ref{eq:E2}) and (\ref{eq:EJ}) that when chemical
potential $\bar{\mu}$ is chosen to be
$\bar{\mu}=3/2B_0-B_0/2-J_1-J_2$ (as given by the stability
condition of $\partial E/\partial \delta_{\ell,i}|_{\rm
var.=0}=0$), linear in $\delta_{\ell i}$ terms vanish. In this
case, only second order terms are left with $E_J$ in (\ref{eq:EJ})
and it justifies the smallness of $E_J$ in accordance with TB.

With (\ref{eq:T})-(\ref{eq:EJ}), one can then derive the
Euler-Lagrange equations of motion for all eight dynamical
variables. They can be linearized and written as
\begin{eqnarray}
\dot{\epsilon}_{\ell,1}^{\prime}&=&2B_0\epsilon_{\ell,1}^{\prime\prime}+
J_1(\epsilon^{\prime\prime}_{\ell,1}-\epsilon^{\prime\prime}_{\ell-1,2})
+J_2(\epsilon^{\prime\prime}_{\ell,1}-\epsilon^{\prime\prime}_{\ell,2}),\nonumber\\
\dot{\epsilon}_{\ell,2}^{\prime}&=&-2B_0\epsilon_{\ell,2}^{\prime\prime}+
J_1(\epsilon^{\prime\prime}_{\ell,2}-\epsilon^{\prime\prime}_{\ell+1,1})
+J_2(\epsilon^{\prime\prime}_{\ell,2}-\epsilon^{\prime\prime}_{\ell,1}),\nonumber\\
\dot{\epsilon}_{\ell,1}^{\prime\prime}&=&-{2\over
B_0}\epsilon_{\ell,1}^{\prime}-2U^\prime B_0\delta_{\ell,1}-
J_1(\epsilon^\prime_{\ell,1}-\epsilon^\prime_{\ell-1,2})\nonumber\\
&&-J_2(\epsilon^\prime_{\ell,1}-\epsilon^\prime_{\ell,2}),\nonumber\\
\dot{\epsilon}_{\ell,2}^{\prime\prime}&=&-{2\over
B_0}\epsilon_{\ell,2}^{\prime}-2U^\prime B_0\delta_{\ell,2}-
J_1(\epsilon^\prime_{\ell,2}-\epsilon^\prime_{\ell+1,1})\nonumber\\
&&-J_2(\epsilon^\prime_{\ell,2}-\epsilon^\prime_{\ell,1}),\nonumber\\
\dot{\delta}_{\ell,1}&=&J_1[2(\nu_{\ell,1}-\nu_{\ell-1,2})-
(\epsilon^{\prime\prime}_{\ell,1}-\epsilon^{\prime\prime}_{\ell-1,2})]\nonumber\\
&&+J_2[2(\nu_{\ell,1}-\nu_{\ell,2})-(\epsilon^{\prime\prime}_{\ell,1}-
\epsilon^{\prime\prime}_{\ell,2})],\nonumber\\
\dot{\delta}_{\ell,2}&=&J_1[2(\nu_{\ell,2}-\nu_{\ell+1,1})-
(\epsilon^{\prime\prime}_{\ell,2}-\epsilon^{\prime\prime}_{\ell+1,1})]\nonumber\\
&&+J_2[2(\nu_{\ell,2}-\nu_{\ell,1})-
(\epsilon^{\prime\prime}_{\ell,2}-\epsilon^{\prime\prime}_{\ell,1})],\nonumber\\
\dot{\nu}_{\ell,1}&=&-(1+3U^\prime)B_0\epsilon_{\ell,1}^\prime-3B_0U^\prime
\delta_{\ell,1}\nonumber\\
&&-{J_1\over 2}[(\delta_{\ell,1}-\delta_{\ell-1,2})+
(\epsilon^{\prime}_{\ell,1}-\epsilon^{\prime}_{\ell-1,2})]\nonumber\\
&&-{J_2\over 2}[(\delta_{\ell,1}-\delta_{\ell,2})+
(\epsilon^{\prime}_{\ell,1}-\epsilon^{\prime}_{\ell,2})],\nonumber\\
\dot{\nu}_{\ell,2}&=&-(1+3U^\prime)B_0\epsilon_{\ell,2}^\prime-3B_0U^\prime
\delta_{\ell,2}\nonumber\\
&&-{J_1\over 2}[(\delta_{\ell,2}-\delta_{\ell+1,1})+
(\epsilon^{\prime}_{\ell,2}-\epsilon^{\prime}_{\ell+1,1})]\nonumber\\
&&-{J_2\over 2}[(\delta_{\ell,2}-\delta_{\ell,1})+
(\epsilon^{\prime}_{\ell,2}-\epsilon^{\prime}_{\ell,1})].
\label{eq:motion}
\end{eqnarray}

\begin{figure}[th]
\vspace{-2.3cm} \mbox{\epsfxsize=1.05\hsize{\epsfbox{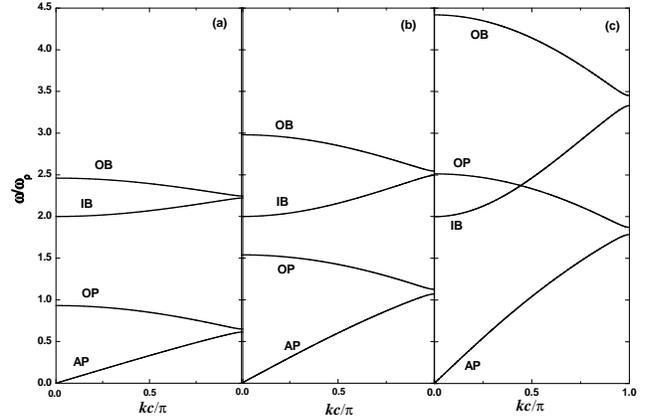}}}
\vspace{-6.1cm} \caption {Dispersions of four-branch modes
resulting from (\ref{eq:motion}). The parameters $J_1=0.09$ and
$J_2=0.1$ are the same for all three frames, while $U^\prime=1$,
$10$, and $100$ respectively for frame (a), (b), and (c). Here IB,
OB, AP, and OP denote for in-phase breathing, out-of-phase
breathing, acoustic phonon, and optical phonon modes. See text for
more description.} \label{fig2}
\end{figure}

\noindent Note that the above eight linear first-order
differential equations can be cast into four second-order ones in
terms of any two sets of variables $\epsilon_{\ell,i}^\prime(t)$,
$\epsilon_{\ell,i}^{\prime\prime}(t)$, $\delta_{\ell,i}(t)$, and
$\nu_{\ell,i}(t)$ ($i=1,2$) only. (That is in the lowest-order
limit, the other two sets will share the same excitation spectra.)
Considering the coupled equations between
$\epsilon_{\ell,i}^\prime(t)$ and $\delta_{\ell,i}(t)$, for
example, and searching for solutions of the type:
$\epsilon_{\ell,i}^\prime(t)\equiv \epsilon_i^\prime e^{i(\ell
kc-\omega t)}$ and $\delta_{\ell,i}(t)\equiv \delta_i e^{i(\ell
kc-\omega t)}$, a $4\times 4$ dynamical matrix (not shown) will be
attained, which is then diagonalized to obtain four branches of
modes. In Fig.~\ref{fig2}, dispersions of the four branches of
modes are shown for three cases: with same $J_1=0.09$ and
$J_2=0.1$, but different $U^\prime=1$, $10$, and $100$. In the
study of Bose-Hubbard model on the SF-Mott insulator quantum phase
transition in 3D, it has been established that $U/J=z\times 5.8$
(with $z$ the number of nearest-neighbor sites) is a critical
point for the case of one atom per site on average ($N=1$). Below
(above) it the system is in the SF (insulating) phase
\cite{oosten01}. Here we are interested in collective excitation
of condensates with average atom number per site being $N\sim
10-100$. Thus the system being in the SF phase is justified when
$U^\prime/J<100$ (recall $U^\prime\sim N U$). Consequently the
$U^\prime=1$ and $10$ cases in Fig.~\ref{fig2} are considered to
be in the SF phase, while the $U^\prime=100$ case may be close to
the phase boundary or in the insulating phase.

The four branches of modes are labelled by in-phase breathing
(IB), out-of-phase breathing (OB), acoustic phonon (AP), and
optical phonon (OP) modes respectively (see Fig.~\ref{fig2}).
Strictly speaking, these descriptions are exact only when phonon
modes are completely decoupled from breathing modes and at the
same time, the propagating wave comprised of
$\epsilon_{\ell,1}^\prime(t)$, for instance, has a phase angle
$\pi$ shift (out-of-phase) by the the propagating wave comprised
of $\epsilon_{\ell,2}^\prime(t)$. The coupling effect, which is
proportional to the magnitude of tunnelling strength $J$, is
indeed small in the TB limit. One should be noted that the
coupling effect is also accompanied by the factor of $B_0
U^\prime$ (recall that $B_0$ corresponds to the size of
condensate), to which it becomes more and more important in the
large $U^\prime$ limit.

At the long-wavelength limit ($kc\ll 1$), dispersions of the four
branches of modes can be analytically solved and uniformly written
as
\begin{eqnarray}
\omega_i^2=\omega_{{\rm 0}i}^2+u_i^2k^2, \label{dispersion}
\end{eqnarray}
where, at the TB limit (keeping to $J$ linear order), one obtains
\begin{eqnarray} \omega_{{\rm 0}i}^2 &=&
  \left\{
  \begin{array}{ll}
   \displaystyle{4+8B_0(J_1+J_2)\left(1+U^\prime+{B_0^2 {U^\prime}^2\over 2}\right)} &
   \displaystyle{i={\rm OB}}
   \\
   \displaystyle{4} & \displaystyle{i={\rm IB}}
  \\
   \displaystyle{8B_0 U^\prime(J_1+J_2)\left(1-{B_0^2 U^\prime\over 2}\right)} &
   \displaystyle{i={\rm OP}}
   \\
   \displaystyle{0} & \displaystyle{i={\rm
   AP}},
   \end{array}
  \right.\nonumber\\
\label{omega0}
\end{eqnarray}
and
\begin{eqnarray} u_{\rm IB}^2 &=& -u_{\rm OB}^2
  = 2B_0(1+U^\prime)c^2 {J_1 J_2\over J_1+J_2} \nonumber\\
   u_{\rm AP}^2 &=& -u_{\rm OP}^2
  = 2B_0 U^\prime c^2 {J_1 J_2\over J_1+J_2}.
\label{vi}
\end{eqnarray}
As shown in (\ref{dispersion})--(\ref{vi}), the slopes ($u_i$) and
the $k=0$ intercept ($\omega_{{\rm 0}i}$) of the four branches of
modes depend crucially on the values of $J_1$, $J_2$, and
$U^\prime$. Therefore observation of these modes will be a direct
way to access them. Some features worth noting are as follows. For
IB mode, the $k=0$ frequency ($\omega = 2 \omega_\rho$) is robust
regardless of the values of $J_1$ and $J_2$. It is a clear result
to the uniform (single-condensate) limit. For AP mode, on the
other hand, the phonon velocity at small $k$ behaves like $u\sim
\sqrt{B_0 U^\prime J}c$, an expected result for the current system
of a bulk modulus $\sim B_0 U^\prime J$. In the large $U^\prime$
case [see Fig.~\ref{fig2}(c)], IB and OP modes start to hybridize,
to which a small gap (not visible in the scale potted) develops
where the two branches cross. This feature may be a signature when
the system undergoes a quantum phase transition from SF into
insulating phase.

In summary, oscillations of Bose condensates in a 1D optical
lattice with a two-point basis is investigated. Focuses are made
to the transverse breathing modes and longitudinal phonon modes.
Pertaining to condensates in site 1 and site 2, in-phase and
out-of-phase modes are obtained. The mode dispersion relations
depend crucially on the sizes of the two intersite Josephson
tunneling strengths $J_1$ and $J_2$ and the on-site repulsion $U$.
For the phonons, there are optical as well as acoustic modes, in a
close resemblance to those in a 1D crystalline solid. While direct
observation of these in-phase and out-of-phase oscillations of
condensates may be resolution limited by current instruments,
dispersion relations of these modes should be probable by
inelastic scattering measurements.

%\acknowledgements

We are grateful to Prof. D.-J. Han for stimulating discussion.
This work is supported by National Science Council of Taiwan under
Grant No. 94-2112-M-003-011.

%\bibliography{bec}
%\bibliographystyle{prsty}

\end{document}